# Clearly Discriminate the Continuum Band and Exciton State of the Hybrid Lead Bromide Perovskite

Jiangjian Shi,[1] Yiming Li,[1,4] Jionghua Wu,[1,4] Huijue Wu,[1] Yanhong Luo,[1,4] Dongmei Li,[1,4] Jacek J. Jasieniak,[3*] Qingbo Meng[1,2*]

1. *Key Laboratory for Renewable Energy, Chinese Academy of Sciences, Beijing Key Laboratory for New Energy Materials and Devices, Institute of Physics, Chinese Academy of Sciences, Beijing 100190, P. R. China.*
2. *Center of Materials Science and Optoelectronics Engineering, University of Chinese Academy of Sciences, Beijing 100049, China*
3. *ARC Centre of Excellence in Exciton Science, Department of Materials Science and Engineering, Monash University, Clayton, Victoria 3800, Australia.*
4. *Department of Physics Science, University of Chinese Academy of Sciences, Beijing 100049, P. R. China*

*Corresponding authors:
*jacek.jasieniak@monash.edu,*
*qbmeng@iphy.ac.cn.*

**Abstract:** Electronic states of the hybrid perovskite enable their promising applications as distinctive optoelectronic materials. The understanding of their electronic structures and charge characters remains highly controversial. The electronic mechanism such as reabsorption, Urbach tail and indirect band for interpreting dual-peak emissions is one of the controversial focuses. Herein, we report that through heterojunction enhanced exciton dissociation and global tracing of multiple radiative electronic states across wide temperature regions, we have succeeded in directly observing free carrier emissions from the hybrid lead bromide perovskite and clearly discriminating the direct continuum band and exciton states. The widely-concerned dual-peak emissions are clarified to be excitonic, arising from two types of exciton states of the perovskite. These excitons possess giant binding energies and superior phase stability compared to conventional inorganic semiconductors, providing important implications for exploiting the excitonic mechanism for realizing novel optoelectronic applications.

**Introduction**

Hybrid lead halide perovskites have emerged as lucrative optoelectronic materials for light-emitting, detecting and photovoltaic applications, owing to their advantages in the bandgap, carrier lifetime and light absorption.[1-12] These advantages have arisen because of the distinctive semiconductor and photophysics properties that such materials possess, including ultrahigh bandedge density of states,[13] ultralow charge trapping cross-section,[14] and low defect densities.[15] Recent findings, including the ultraslow hot carrier cooling,[16-17] ultralong polarization memory,[18] giant spin-orbit Rashba splitting[19] and weak electron-lattice coupling,[20] further increase mystery of the perovskites. Underlying these alluring characteristics are intrinsic electronic properties of the halide perovskite materials, while electronic structure ultimately determines their photoelectric and energy conversion mechanisms and performances.[21-22] As such, exploring and exploiting electronic states of the hybrid perovskite would help improve performance of current devices and expand research and application horizons of these materials.

Although a variety of efforts were paid, a clear understanding of the intrinsic electronic structure and charge-carrier characters of the perovskite remains lacking, and controversies on these properties have always been ongoing. These endless controversies arise because of the observed complicated spectroscopic characteristics and distinct scenarios proposed for interpreting them. The origin for dual-peak emissions within both the single crystal and polycrystalline perovskites is one of the controversial focuses.[23-30] Reabsorption provided a satisfied explanation for this phenomenon within the single crystal methylammonium lead bromide (MAPbBr$_3$) where the light scattered from the crystal edges produces the second emission peak;[23] however, this cannot account for the MAPbBr$_3$ thin film because light reabsorption and edge scattering are negligible here.[20] Indirect band structure induced by the static or dynamical Rashba effect arising from the strong spin-orbit coupling and symmetry breaking is another common interpretation[28-30] since a giant Rashba splitting within MAPbBr$_3$ was indeed observed using angular resolved photoelectron spectroscopy.[19] The indirect band structure scenario was also proposed to explain an observed unusual temperature-dependent second-order radiative recombination velocity.[31] However, this needs to be carefully rechecked because a different experimental result of the radiative recombination velocity was observed recently.[32] Moreover, whether the Rashba effect and its relevant theoretical prediction can indeed be reflected by the photophysics characteristics in conventional experiments is still an open question. The dual-peak emission was also attributed to the exciton states of the perovskite;[20] however, this phenomenon was recently claimed to arise from free carriers of the $MAPbBr_3$.[29] Essentially, these controversies regarding the dual-peak emission are debates and explorations on the bandedge electronic structure and charge nature (i.e. direct-indirect band and exciton-free carrier characters) of hybrid perovskites.

Herein, we provide a new and important experimental finding to clarify the controversial bandedge electronic structure of the $MAPbBr_3$ perovskite. By enhancing the exciton dissociation, we realized an observation of higher-energy emissions arising from free carriers that populated within the direct continuum bands. These emissions are clearly distinct from the exciton dual emissions in the energy space and exhibit little signature of indirect states. Comparisons of direct bandgap and exciton energy level derived from light absorption and photoluminescence (PL) of the $MAPbBr_3$, respectively, make a confirmation of this energy landscape. A bound exciton state was further determined according to its correlation to the traps of the perovskite. These findings help to conclude that (1) the commonly observed dual-peak emission of the $MAPbBr_3$ is arisen from the dual exciton states and not the free carriers of the perovskite; (2) the free carrier emission is arisen from the direct continuum band; and (3) excitons dominate the charge species and photophysics characteristics of the $MAPbBr_3$ even at room temperature. Therefore, this work draws a clear picture of fine structure of the multiple electronic states within the $MAPbBr_3$ including continuum band, free (FE) and bound excitons (BE), and trap states, and provides fresh insight into clearly understanding the electronic structure and photophysics mechanisms of the hybrid perovskite. Moreover, these findings present important opportunities of exploiting tunable electronic states and especially the room-temperature stabilized exciton states for realizing more abundant optoelectronic applications of the hybrid perovskite materials.

**Results**

Dual-peak emissions within the hybrid perovskites including both the $MAPbI_3$ and $MAPbBr_3$

have been widely concerned.[23-30] MAPbBr$_3$ was usually chosen for further photophysics mechanism studies because (1) it gives emissions in the visible regions which is much easier to be optically probed and (2) it possesses more stable crystal phase and spectroscopic structures across wide temperature regions. Figure 1(a) presents typical PL spectra of the MAPbBr$_3$ film at room temperature at different decay times after the photoexcitation. Shoulders in the lower-energy regime can be unambiguously observed. The PL spectra probed at 1 ns is extracted for further emission peak fittings. Clearly, these asymmetric spectra are composed of two emission bands centered at ~2.32 eV and 2.24 eV, respectively, which is denoted as peak 1 and peak 2 for clarity here and in the following will be demonstrated to originate from FE and BE states, respectively. Due to Stokes shifts arising from phonon involvement or tail state, these two emission bands are located obviously below both the direct bandgap ($E_g$) and the exciton energy level of the MAPbBr$_3$, which makes it difficult to determine their photophysics mechanism by direct energy level comparisons. The reabsorption mechanism once proposed can be firstly excluded here by using a reflection PL configuration and eliminating the edge light scattering. As such, this dual-peak emission behavior possesses a direct correlation to bandedge electronic structures of the MAPbBr$_3$.

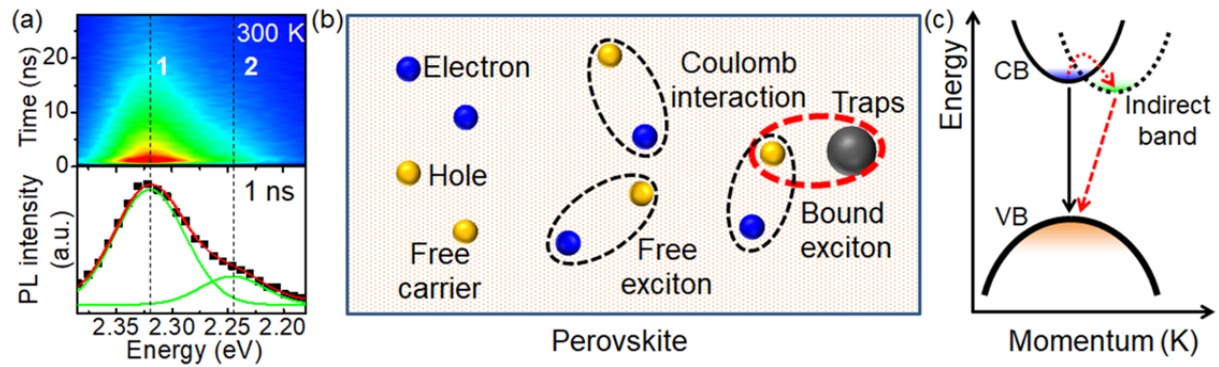

**Figure 1. Dual-peak emissions of the MAPbBr$_3$.** (a) Typical emission spectra of the MAPbBr$_3$ film at room temperature. The dual peaks are decoupled by using multiple-Gaussian fittings. Possible charge characters and energy band structure to interpret the dual-peak emission, including (b) mixed charge species within the perovskite (e.g. free carrier, free and bound excitons) and (c) indirect band structure of the perovskite. Charge transfer can occur between the conduction band (CB), indirect band and the valence band (VB).

Figure 1(b-c) presents a summary of current understandings of charge species and energy band structure of the perovskite to account for the dual-peak emission. It was generally believed that free carrier dominates the charge species and thus the radiative recombination (photon emissions) of the perovskite.[29, 33] Excitons including FE and BE may also exist within the perovskite, influencing the photophysics characteristics.[20,34] For free carriers, it was proposed that indirect band structure could influence their distribution properties and result in dual-peak emissions.[28-31] This indirect band and its relevant local electric field were believed to be able to prolong the carrier lifetime and enhance exciton dissociation, beneficial for photovoltaic or photodetector devices, however stifling the light emitting and optical gain performances.[30] Nonetheless, all of these understandings are still in controversial.

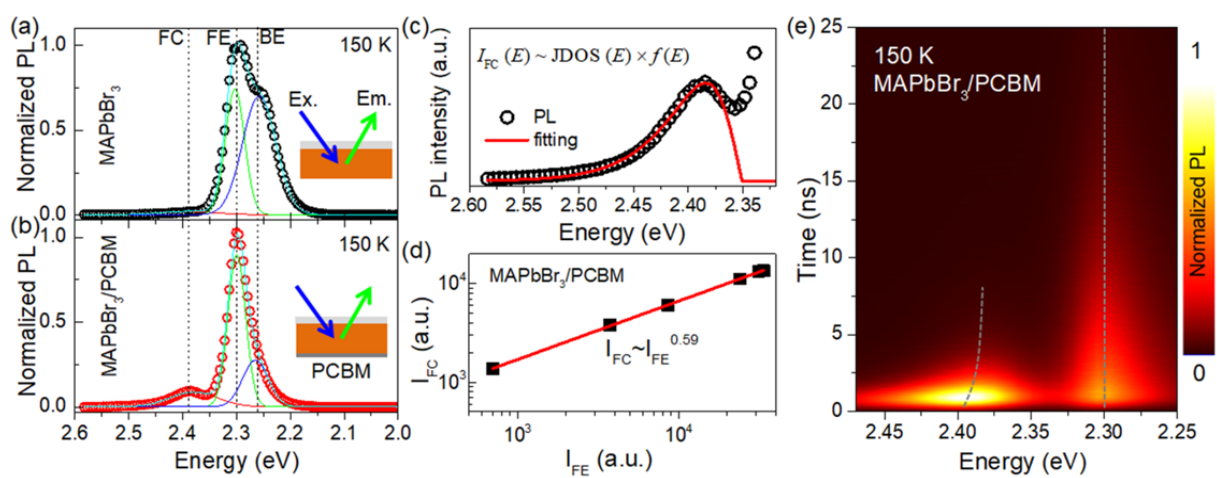

**Figure 2. A direct observation of the continuum band emission of MAPbBr$_3$.** PL spectra of the perovskite MAPbBr$_3$ films, and their multiple-peak fittings (a. without and b. with a PCBM layer on the surface). The insets depict the PL measurement configuration. (c) Fitting of the continuum band emission spectrum according to the carrier distribution derived from joint density of states (JDOS) of the perovskite and the Fermi-Dirac function. (d) An experimental relationship between the free exciton emission intensity ($I_{FE}$) and the free carrier emission intensity ($I_{FC}$) of the MAPbBr$_3$/PCBM sample. This relationship can be well fitted using a power-law equation. (e) Two-dimensional pseudocolor plots of transient PL for the MAPbBr$_3$/PCBM film as functions of emission energy and time. The dashed lines depict evolutions of the FC and FE emission peak position with time.

These controversies arise because that the dual-peak emission spectra of the perovskite only provide limited information about the energy structure, and that diversified electron-lattice properties of the perovskite can complicate the emission characteristics. Nonetheless, if we could observe optical response, such as PL emission, from the non-prominent electronic states, it will then be possible to determine the mechanisms for the current spectroscopic structure. Under the guidance of this idea, we introduced an organic [6,6]-phenyl-C61-butyric acid methyl ester (PCBM) layer onto surface of the $MAPbBr_3$ film to construct a $MAPbBr_3$/PCBM heterojunction.[35] The electric field within the bulk perovskite induced by this heterojunction would significantly enhance the exciton dissociation and the proportion of free carriers, and thus modulate the spectroscopic features. To avoid the interface influence of the PCBM layer, we probed PL spectra of the perovskite from the opposite side, that is, from the quartz substrate side, and used an excitation wavelength of 445 nm (~2.8 eV) to obtain a short optical penetration depth of <100 nm. The pump fluence of the excitation pulse is ~1 nJ $cm^{-2}$, which is low enough to avoid nonlinear effects and multi-particle scatterings.

Surprisingly, a new emission band located at the higher-energy side to the original dual-peak emission appeared. This phenomenon can always be observed across the whole studied temperature range from 10 to 300K (Supplementary Figure S1). In the meanwhile, the intensity of the original dual-peak emissions and especially the lower-energy emission band (BE) is obviously weakened. Figure 2(a-b) present the PL spectra of the $MAPbBr_3$ and $MAPbBr_3$/PCBM samples at 150K, where the emission bands are qualitatively decoupled using multiple-Gaussian fittings. Specially, the above-mentioned dual-peak emissions located at ~2.3 eV and ~2.26 eV, respectively, can be observed in both samples. For the $MAPbBr_3$/PCBM sample, a new asymmetric emission band appears at ~2.38 eV, which is denoted as FC for clarity here and we will confirm its emission mechanism in the following. Notably, this new emission band should arise from the bulk region of the $MAPbBr_3$ thin film because the PL emission was collected from the quartz side with a reflection configuration (as depicted in the inset of Figure 2(a-b)) and the PCBM layer would not influence the quartz/$MAPbBr_3$ interface. The non-local electric field and charge redistribution within the $MAPbBr_3$ caused by the PCBM heterojunction can be the origin for this result. Clearly, both the electric field and charge redistribution can weaken the stability of the exciton.[21-22] Thus, we infer that the original dual-peak emission is excitonic and the new emission band is originated from the free carriers populated within the continuum bands.

We take a closer look at this new emission band, as presented in Figure 2(c). This asymmetric emission band exhibits a broad extending toward the high-energy side and seems possess a relatively sharp cutoff edge at the low energy side. This spectral signature is similar to the free carrier emissions from continuum bands.[22] To demonstrate this scenario, we firstly attempted to make a fitting of this energy ($E$)-dispersed emission band ($I_{FC}(E)$) according to the carrier distribution within the continuum bands, which can be described using the joint density of states (JDOS) and the Fermi-Dirac function ($f(E)$), that is,[22, 32]

$$I_{FC}(E) \sim JDOS(E) \times f(E)$$
$$\sim (E - E_g)^{1/2} \times \left[1 + \exp\left(\frac{E - E_f}{K_B T}\right)\right]^{-1}, \quad (1)$$

where $E_f$ is the quasi Fermi energy level, $K_B$ is the Boltzmann constant and $T$ is the absolute temperature. Further considering the broadening of continuum bands,[36] equation (1) provides a satisfied fitting of this emission band in energy region from 2.37 eV to 2.6 eV, thereby supporting the reasonability of attributing this new emission band to free carriers.

More evidence can be obtained by studying the experimental relationship between the FE emission intensity ($I_{FE}$) and the free carrier emission intensity ($I_{FC}$) of the MAPbBr$_3$/PCBM sample under different excitation intensities. As in Figure 2(d), the $I_{FC}$ exhibits approximately half-power dependence to the $I_{FE}$, while the $I_{FC}$ is linear to the excitation intensity. This result may be beyond our expectation because it is usually thought that the $I_{FC}$ possesses quadratic dependence to the excitation intensity while the $I_{FE}$ exhibits a linear relationship.[29-30] Here, it should be noted that this previous idea is valid only when studying the relationship between the transient PL peak at the initial time after photoexcitation and the excitation intensity and without significant charge transfer, such as charge trapping or exciton dissociation.[37-38] If the materials under different excitation intensities possess a same emission mechanism and without any significant change in PL quantum yields, the steady-state or time integrated PL intensity of all the emission bands, no matter the free carrier or the exciton, should follow a linear dependence to the excitation intensity. This is linear optics. Thus, previous works that using the quadratic or super-quadratic relationships between the PL intensity and the excitation intensity to determine the free carrier mechanism of the above-mentioned dual-peak emissions may not be reliable.

The experimental relationship of $I_{FC} \sim I_{FE}^{0.59}$ we observed here can be well interpreted by using the Saha-Langmuir theory which describes the population balance between the free carrier and the exciton within one semiconductor system. Under low excitation intensities, the free carrier density is approximately linear to while the free exciton density is quadratic to the excitation intensity (Supplementary Figure S2). Under high excitation intensities, the free carrier density is half-power to and the free exciton density is linear to the excitation intensity.[39] Therefore, free carrier density is always exhibits a half-power relationship to the free exciton density, which consequently result in a relationship of $I_{FC} \sim I_{FE}^{1/2}$. This agrees well with our experimental result, thereby further supporting that the new emission band is arisen from free carriers and the original dual-peak emissions are excitonic.

We further measured two-dimensional PL spectra at different decay times to observe the dynamics behavior of these emission bands (Figure 2(e)). Under the time resolution of our transient PL measurement, the FE and FC emissions rise to their peaks simultaneously, at ~ 1.5 ns after the photoexcitation. This time is significantly shorter than the charge transport time of the perovskite,[12] which helps us to completely exclude the possibility that new emission band is induced by the interface states of the $MAPbBr_3$/PCBM heterojunction. Clearly, the FC decay lifetime is much shorter than that of the excitons, confirming their different charge recombination or transport mechanisms. More interestingly, the FC emission exhibits a slight red-shift with time while the peak position of the exciton emissions remains unchanged across the whole time range. This arises because the Fermion nature determines that free carriers would relocated toward lower-energy states when their density is reduced with time due to recombination or interface extraction while the energy level and distribution of the exciton (Boson) is almost independent to their density.[21-22] Gathering the above results together, we can conclude that it is reasonable to attribute the original dual-peak emissions and the new observed higher-energy emission band to excitons and free carriers, respectively.

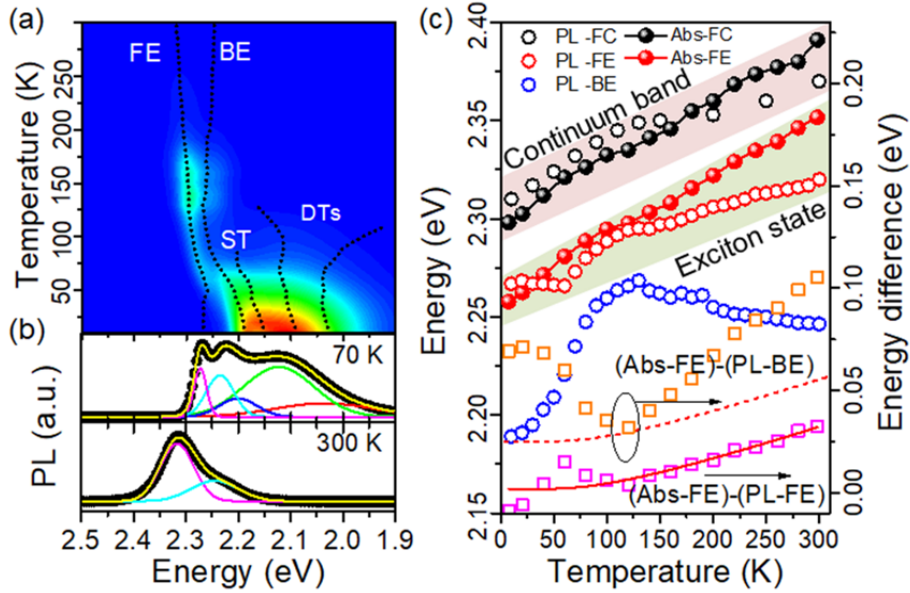

**Figure 3. Distinguish between continuum band and exciton state emissions.** (a) Two-dimensional pseudocolor plots of steady-state PL spectra for the MAPbBr$_3$ film as functions of emission energy and temperature. The emission bands including FE, BE, shallow trap (ST) and deep traps (DTs) across the whole temperature range are decoupled from each other, as indicated by the dashed lines. (b) Peaks fittings of emission spectra at 70K and 300K, respectively. (c) Comparisons of direct bandgap and exciton energy level of the MAPbBr$_3$ derived from light absorption and PL, respectively.

To confirm our findings, a direct comparison of the emission energy positions to the energy levels derived from light absorption of the MAPbBr$_3$ was made. Because of the existence of multiple radiative electronic states within the MAPbBr$_3$,[20] we made a global tracing of the emission bands across the whole temperature range from 10K to 300K. The spectral features and energy position of each band were successfully decoupled by following several rules (Supplementary Figure S3), including (1) spectra residual should be the least under multi-Gaussian-peaks fittings, (2) emission energy position, intensity and full width at half maximum (FWHM) of adjacent temperatures should be continuous, and (3) emission bands of different samples should be similar to each other and without any untraceable mutation. Under these rules, up to five emission bands have been found within the MAPbBr$_3$ sample at temperatures lower than 100K (Figure 3(a)). These emission bands can be assigned to FE, BE, shallow and deep traps, respectively, where the trap states will be further studied elsewhere.

Figure 3(b) gives two examples of the emission bands fittings at 70K and 300K, respectively. After the orthorhombic-tetragonal phase transition, these trap emissions cannot be observed anymore, while both the FE and BE emissions are retained up to the room temperature. The FC emission spectra across the whole temperature range can always be fitted by equation (1). Figure 3(c) summarizes the energy levels derived from the light absorption (reused from a previous work[36]) and PL spectra of the $MAPbBr_3$, respectively. Clearly, the $E_g$ derived from the PL (PL-FC) is always close to the $E_g$ from the light absorption (Abs-FC) and energy level of the FE derived from the PL (PL-FE) is close to that from the light absorption (Abs-FE), especially at temperatures lower than 150K. These results confirm that the widely-concerned dual-peak emissions are indeed arisen from the exciton of the $MAPbBr_3$. The energy difference ($\Delta E$) between the PL-FE and Abs-FE at high temperatures may come from the phonon effect because it can be well fitted by using a temperature-dependent phonon occupation model,[30] that is, $\Delta E=\Delta E_0+A/(\exp(E_{ph}/K_BT)-1)$, where $A$ is a prefactor, $E_{ph}$ is the photon energy and $\Delta E_0$ is a temperature-independent term. The $E_{ph}$ is derived to be ~26 meV, almost same to the longitudinal optical phonon (LO) energy that derived from the temperature-dependent FWHM of the FE emission (Supplementary Figure S4). This photon energy is also very close to the value of the $\Delta E$ at room temperature, indicating that the FE emission is a FE-LO process.

The energy position of the BE emission peak is also presented in Figure 3(c), which exhibits a non-monotonous evolution with temperature. This evolution relationship does not agree with the theoretical prediction using the above phonon occupation model (dashed line in Figure 3(c)). The giant $\Delta E$ of >70 meV at both room temperature and low temperatures between this emission position and the FE energy level is far beyond the phonon energy of the perovskite.[40] Therefore, the BE emission here should not be a phonon replica of the FE state. In addition, the giant $\Delta E$ excludes the possibility of biexciton and trion (Supplementary Table 1).[21] Clearly, this emission possesses dependence to the crystal phase of the perovskite and exhibits a turning due to the phase transition, which is a little similar to the transition behavior of the trap state within the $MAPbBr_3$ as we once reported.[20] We have also previously shown that a certain level of trap density is necessary for producing this BE emission.[20] As such, we infer that this emission is closely correlated to the trap state of the perovskite, implying the possibility of bound exciton.

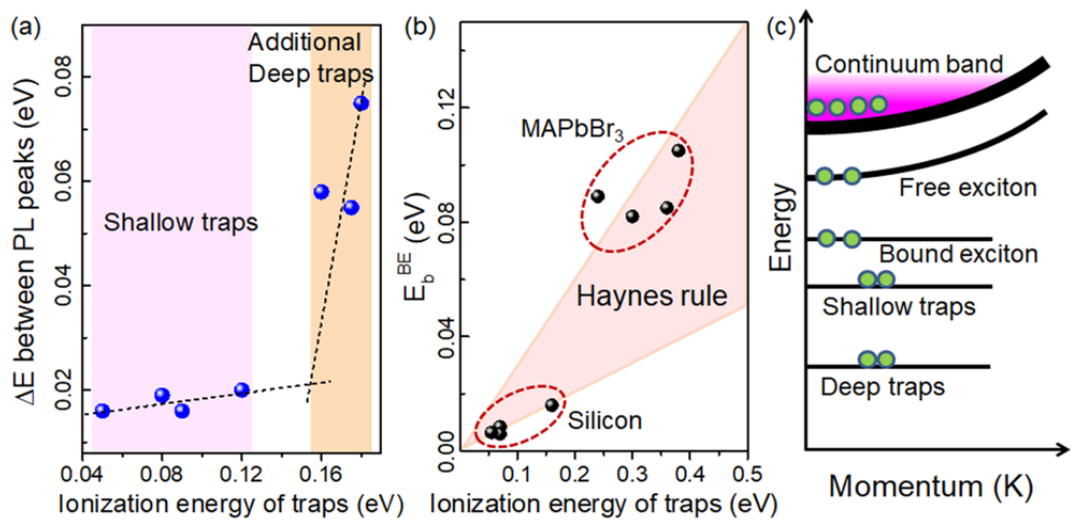

**Figure 4. Determination of bound exciton (BE) within the MAPbBr$_3$.** (a) An experimental relationship between the ionization energy of trap states and energy difference ($\Delta E$) of the PL dual emission peaks. The dashed lines are guides for eyes, illustrating a sharp increase in the $\Delta E$ when additional deep traps are introduced. (b) A direct comparison of binding energies of the BE state ($E_b^{BE}$) between the perovskite and the conventional silicon within the Haynes rule framework. (c) Structures of multiple energy states that participated in the photoelectric conversion processes of the MAPbBr$_3$ perovskite.

Bound exciton is formed by the Coulomb interaction between a free exciton and a trap center,[21] as depicted in Figure 1(b), with binding energies ($E_b^{BE}$) correlated to the ionization energy of the traps. By controlling the crystallization of the MAPbBr$_3$ films, we can tune their trap state properties in experiment.[41] Figure 4(a) presents the relationship between $\Delta E$ of the PL emission dual peaks and the ionization energy of traps determined from the thermal admittance spectroscopy (Supplementary Figure S5).[42] Two types of trap states were observed within these samples, that is, shallow traps with ionization energies ($E_I^{ST}$) ranging from 0.05 eV to 0.18 eV and deep traps with ionization energies ($E_I^{DT}$) larger than 0.2 eV. The shallow traps exist within all the perovskite films while the deep traps only appear under certain perovskite crystallization conditions. For these MAPbBr$_3$ who only possess shallow traps, their $\Delta E$ is lower than 0.02 eV and exhibits a slight dependence to the $E_I^{ST}$; surprisingly, the $\Delta E$ is significantly increased to >0.05 eV and beyond the original dependence between $\Delta E$ and the $E_\text{I}^\text{ST}$ when additional deep traps are introduced. This means that the giant $\Delta E$ of the dual-peak emission of the MAPbBr$_3$ is induced by the deep traps, which is a clear signature of BE states. A comparison of the $E_\text{b}^\text{BE}$ to the $E_\text{I}^\text{DT}$ within the Haynes rule framework[21] provides another support for the BE state scenario, where the $E_\text{b}^\text{BE}$ is calculated from the energy difference between the BE emission position and energy level of the free exciton. A quasi linear relationship exists between the $E_\text{b}^\text{BE}$ and the $E_\text{I}^\text{DT}$ (Figure 4(b)), agreeing well with the Haynes rule that the $E_\text{b}^\text{BE}$ is usually in the range between 0.1 $E_\text{I}^\text{DT}$ and 0.3 $E_\text{I}^\text{DT}$.[21] Therefore, we can conclude that we have clearly discriminated the continuum band and exciton states of the perovskite MAPbBr$_3$. The dual-peak emissions should be excitonic; specially, the second peak arises from the BE state.

In previous works, this second emission band was ascribed to the free carriers populated within indirect bands induced by static or dynamical Rashba effect.[29-30] This interpretation may be attractive because it can enhance the mystery of the perovskite materials. However, completely reliable experimental and theoretical results remain lacking. Firstly, as discussed above, the dependence between the emission intensity and excitation intensity can hardly be used to determine the emission mechanism if significant charge transfer occurs within the material system. It has been widely demonstrated that rapid charge transfer exists between multiple electronic states of both the bromide and the iodide perovskites with an orthorhombic phase at low temperatures.[20, 29] Secondly, most of the experimental results about the temperature-dependent second-order radiative recombination velocity do not agree with the indirect band model but strongly suggest that the emission is from direct band.[32] Thirdly, it is inherently unreasonable to attribute the emission band of the bromide perovskite at low temperatures such as 50K or 77K to the free carriers because the charge species are dominated by excitons at such low temperatures according to the Saha-Langmuir theory (Supplementary Figure S2). Based on the above three major reasons, more other details in previous works and our own experimental results presented above, we prefer to doubt the reasonability of the indirect band (or indirect tail band) scenario in interpreting the photoelectric characteristics of the perovskite. Nonetheless, whether an indirect band structure or Rashba effect really exist and influence other physics properties of the perovskite is still an open question.

**Discussions**

Based on the above determined emission mechanisms, we have established a clear physics picture of the electronic structure that directly participating in the photoelectric processes of the MAPbBr$_3$ (Figure 4(c)). Below the direct continuum band, MAPbBr$_3$ possesses two types of exciton states (i.e. FE and BE), resulting in the widely-concerned dual-peak emissions. The deep trap states within the MAPbBr$_3$ contributes to formations of the BE state. These findings provide important experimental evidences for eliminating controversies on energy structure of the perovskite, and shed new insight into the charge-carrier character of these materials. Moreover, the confirmed exciton states here bring endless opportunities for excitonic applications of the perovskite materials.

The MAPbBr$_3$ possesses an exciton binding energy of ~40 meV, which is much higher than that of inorganic semiconductors with a similar $E_g$, such as ZnTe (~12 meV) and CdS (~25 meV), as well as wider $E_g$ candidates, such as GaN (~25 meV) and ZnS (~28 meV).[21] Firstly, this large binding energy helps to stabilize the excitonic phase of the perovskite at relatively high temperatures. In our observations here, excitons still dominate the charge occupation and recombination properties of the MAPbBr$_3$ at room temperature. This mainly benefits from an ordered lattice and bandedge structure of the perovskite, which exhibits little band fluctuation and tail states.[27] In fact, we have observed that excitonic effect of the perovskite can be significantly weakened when increasing the band fluctuation by disordering the film crystallization.[41] A weak electron-lattice coupling strength of the lead-halide perovskites may also lead to large charge transition energy barriers between the excitons and free carriers, and thus contribute to the stabilization of exciton states.[20] In addition, possible couplings between exciton and other quasi-particles, such as forming exciton polariton,[21] may also help to stabilize the excitonic phase of the MAPbBr$_3$ at room temperature. Secondly, this large binding energy leads to a high Mott density of ~5×10$^{17}$ cm$^{-3}$ at room temperature.[41] This helps to stabilize the excitonic phase at high photoexcitation densities, which provides a sufficient space for realizing population inversions of the exciton state and especially the BE state. In our other experiments, optical amplification of the BE spontaneous emission have

been realized, which also further support the reliability of our above results.[41]

The $E_b^{BE}$ of ~100 meV of the MAPbBr$_3$ is among the largest reported for bulk semiconductors,[20] which has approached the upper limit of Haynes rule, as presented by a direct comparison between the MAPbBr$_3$ and the silicon.[43] It is evident that what enables such optoelectronic properties to emerge within perovskites is the underlying presence of tunable defects.[44] The relatively wide and stable phase regime of this ternary semiconductor provides a large parameter space for self-doping and trap-state engineering using simple solution processes. These BE states located much below the bandedge can help extend the light absorption edge of the perovskite; while the strong electric field within the perovskite devices make these BE charges possible to delocalization and to contribute to photocurrent enhancement. With exciton states being sensitive to external fields, such as electric or magnetic, high sensitive electrical-optical and magnetic-optical coupled devices can be expected for based on the hybrid perovskite material. Moreover, the spin properties of the exciton state provide implications for quantum manipulation and communication.[45] The highly stabilized exciton nature of the perovskite helps to make these applications practical at high temperatures, possibly even at room temperature. Each of these exciting prospects requires a greater level of understanding into the nature of the carrier and exciton states within perovskite materials.

In summary, through heterojunction enhanced exciton dissociation and global tracing of the multiple radiative electronics states across wide temperature regions, we have succeeded in observing the free carrier emission from the MAPbBr$_3$ and clearly discriminating the direct continuum band and exciton states of the hybrid perovskite materials. Our findings show that direct and not the indirect band results in the free carrier emission; and more importantly excitons and not free carriers dominate the emission and photophysics characteristics of the MAPbBr$_3$ even at room temperature. The widely-concerned and controversial dual-peak emissions have been clarified to arise from two types of exciton states, that is, FE and BE states. These two excitons possess giant binding energies and superior phase stability compared to conventional inorganic semiconductors, which provides important implications for exploiting the excitonic mechanism for realizing novel optoelectronic applications, such as extending the light absorption for enhancing photoresponse of perovskite solar cell and photodetector, achieving low-threshold and high-gain perovskite excitonic stimulated emission, and developing electrical-magnetic-optical coupled and even quantum manipulation devices.

**Methods**

*Perovskite film deposition and characterizations.* The MAPbBr$_3$ samples were deposited by using an anti-solvent method, where chlorobenzene is used as the anti-solvent. Briefly, a 1M MAPbBr$_3$ (or MAPbI$_3$) precursor solution in dimethylformamide (DMF) was firstly spread onto a quartz glass/compact Al$_2$O$_3$ substrate. A few of certain seconds after the beginning of spin coating at 4000 rpm, the chlorobenzene was dropping onto the film to produce a fast crystallisation of the perovskite. To adjust the film deposition process, the addition time for applying the anti-solvent was controlled as described in the main text. After 30 s of spinning, the film was transferred to a hot plate at 100 °C and annealed for 10 min. A 50 mg/ml PMMA solution was finally spin coated onto the perovskite film for protection. The phase of the final perovskite was characterised by a Panalytical X-ray diffractometer. The averaged thickness of each film is measured by using a step profiler.

*Optical characterization.* The light transmission and reflectance of the perovskite film were measured with a UV-vis spectrometer (Shimadzu 3600). The light absorption is calculated from the transmission and reflectance. The steady-state and transient PL of the perovskite film were measured using an Edinburgh Fluorescence Spectrometer (FLS 920). A 445 nm pulsed diode laser (EPL-445, ~5 nJ cm$^{-2}$, 62 ps) was used as the excitation source. A circular adjustable neutral density filter was adopted to adjust the excitation intensity. The PL was collected in the reflectance mode, where a PMT together with a TCSPC module was applied to detect the time-resolved or time-integrated PL. For temperature dependent measurements, an ARS liquid helium cryostat was employed as the sample chamber under vacuum, where the temperature was controlled using a Lakeshore controller. The perovskite film on the glass substrate was directly fixed to the optical sample holder by screws. During the temperature adjustment, the system was firstly stabilised for at least 5 min at each temperature. Before the PL measurements, the perovskite sample was always kept in the dark without any bias light or laser illumination to avoid the possible production of charge accumulation and a local electric field.

For the femtosecond transient absorption measurement, a high energy pulse generated from the femtosecond amplifier and the OPA were used as the pump light, and a white light generated by focusing the 800 nm (35 fs) pulse onto a 2 mm-thick sapphire and filtered by a short-pass filter (775 nm, Sigmakoki) was used as the probe light. The probe light was vertically focused onto the sample through a series of achromatic lens, with a final probe spot diameter of <1 mm. The pump light was focused onto the same centre by a concave mirror, with a spot diameter of ~5 mm. There was an angle of ~20° between the pump and probe light, that is, non-degenerate configuration. The time delay between these two pulses was controlled by using an Electric stage (Zolix) together with a hollow retroreflector (Newport). The transmitted probe light was detected by a CMOS optical fibre spectrometer (Avantes, AvaSpec-ULS2048CL-EVO) with a single-pulse mode (integrating time 30 μs). A pinhole 10 cm away from the sample is put in this light path to eliminate the light scattering from pump light and PL. With multiple spectra recording and calculating, the noise of the Δ$A$ is lower than 0.1 mOD. The chirp of the probe pulse was calibrated by measuring the Kerr effect of $CS_2$. The VSL measurement was performed based on another CCD array optical fibre spectrometer (Ocean optics QE pro).

*Electrical characterization.* The trap states of the $MAPbBr_3$ film were measured by using thermal admittance spectroscopy (TAS). The frequency-dependent capacitances of the FTO/compact $TiO_2$/$MAPbBr_3$/Spiro/Au cells were measured by using an electrochemical station (Princeton). The cell was placed inside a vacuum chamber of a low-temperature probe station (Lakeshore TTPX) with liquid nitrogen as the cooling source. The bandwidth of the electrical probe was 1 GHz. For capacitance measurement, the cell was always measured under dark conditions with no bias voltage being applied to the cell. The data analysis was performed according to published literature describing TAS. The trap state energy levels were derived from the Arrhenius curves. Mott-Schottky curves were also measured to help obtain trap density approximations.

**Acknowledgements**

This work was supported by Natural Science Foundation of China (No. 11874402, 91733301, 51761145042, 51421002, 51627803, 51872321) and the International Partnership Program of Chinese Academy of Sciences (No. 112111KYSB20170089). JJ would like to acknowledge funding through the ARC Centre of Excellence in Exciton Science (CE170100026).

**Competing financial interests**

The authors declare no competing financial interests.

**Author contributions**

J. Shi designed this study, and made experimental measurements, theoretical calculations and analysis; Y. Li prepared the sample and made photoluminescence measurements; J. Wu assisted in capacitance and transient absorption measurements and data extraction; H. Wu, Y. Luo and D. Li participated in organizing the measurement setups and assisted in the sample preparation; J. Jasieniak conceived and supported this project and participated in experimental and theoretical analysis; Q. Meng conceived and supported this project. All of authors contributed to the discussion. The manuscript was written by J. Shi, corrected by J. Jasieniak and Q. Meng and approved by all contributions.

**Additional information**

Supplementary information is available in the online version of the paper.